\documentclass[11pt,twoside]{article}
\usepackage{amsmath,amsfonts,amssymb}
\usepackage{latexsym}
\usepackage{dcolumn}
\usepackage{graphicx,epsfig}
\usepackage{amsthm}
\usepackage{color}
\title{\bf  More on the initial singularity problem in gravity's rainbow cosmology }
\author{M. Khodadi$^{1}$\thanks{email: m.khodadi@stu.umz.ac.ir}\, , K. Nozari$^{1}$
\thanks{email: knozari@umz.ac.ir}\, and H. R. Sepangi $^{2}$\thanks{email:
hr-sepangi@sbu.ac.ir}
\\{\small $^1$ \emph{Department of Physics, Faculty of Basic Sciences,
University of Mazandaran,}}\\
{\small {\it P. O. Box 47416-95447, Babolsar, Iran}}
\\{\small $^2$\emph{Department of Physics, Shahid Beheshti University, G.C., Evin, Tehran 19839,
Iran}}}
\begin{document}
\maketitle
\begin{abstract}
Using a one-dimensional minisuperspace model with a dimensionless ratio
$\frac{E}{E_{Pl}}$, we study the initial singularity problem at the
quantum level for the closed rainbow cosmology with a homogeneous, isotropic
classical space-time background. We derive the classical Hamiltonian
within the framework of Schutz's formalism for an ideal fluid with a cosmological
constant. We characterize the behavior of the system at the early stages of
the universe evolution through analyzing the relevant shapes for the potential
sector of the classical Hamiltonian for various matter sources, each separately
modified by two rainbow functions. We show that for both rainbow universe models
presented here, there is the possibility of eliminating the initial singularity by
forming a potential barrier and static universe for a non-zero value of the scale
factor. We investigate their quantum stability and show that for an energy-dependent
space-time geometry with energies comparable with the Planck energy, the non-zero
value of the scale factor may be stable. It is shown that under certain constraints
the rainbow universe model filled with an exotic matter as a domain wall fluid plus
a cosmological constant can result in a non-singular harmonic universe. In addition,
we demonstrate that the harmonically oscillating universe with
respect to the scale factor is sensitive to $\frac{E}{E_{Pl}}$ and that
at high energies it may become stable quantum mechanically. Through a
Schr\"{o}dinger-Wheeler-De Witt (SWD) equation obtained from the quantization of
the classical Hamiltonian, we also extract the wave packet
of the universe with a focus on the early stages of the evolution. The resulting wave packet
supports the existence of a bouncing non-singular universe within the context of
gravity's rainbow proposal.\\
\vspace{0.8mm}\newline \textbf{Keywords}: Rainbow Cosmology; Potential Barrier;
Static Universe; Quantum Stability; Wave Packet
\vspace{1mm}\newline \textbf{PACS numbers}: 04.60.-m; 98.80.Qc; 04.20.Dw
\end{abstract}
\section{Introduction}
One of the serious problems from which standard cosmology suffers is the so-called
``initial singularity problem.'' At first, Einstein and  many others
believed that the origin of this issue goes back to the
simplifying idea of the ``cosmological principle.'' However, this  was challenged by
Lemaitre in \cite{Geo}. He argued that in a certain class of anisotropic
universe models, the tendency towards the appearance of singularities is even greater
than in isotropic ones and concluded that this problem cannot be linked to the cosmological
principle. Later on, Penrose and Hawking  presented some theorems on the
existence of singularities in the solutions of Einstein's field equations \cite{Pen, SW0, SW1, SW2}.
The Penrose theorem considers space-like singularities which are  characteristic of non-rotating
uncharged black-holes, whereas Hawking's singularity theorem covers the whole universe. For a
comprehensive review of the concept of singularity see \cite{PH, J}.

To study the early universe for which quantum gravitational effects are to be considered, General
Relativity (GR) alone is insufficient and quantum gravity (QG) should come to the fore. More precisely,
QG proposal could be used to avoid the initial singularity through a potential barrier \cite{A0}-\cite{A5}.
Theories like Loop Quantum Gravity (LQG) \cite{LQ0, LQ1} and Super Strings (SS) \cite{St} are attempts in
this direction \cite{Ma0, Ma1, Ma2, Ben}.  Despite the absence of a fully self-consistent theory for QG, semi-classical
approaches \footnote{Semi-classical approach here means a theoretical framework in which one treats
matter fields as being quantum and the gravitational field as being classical. Indeed, the matter
fields are propagating on the classical space-time background, as described in GR.}, have been very much
in the spotlight in recent years. Generally, semi-classical QG approaches contain two eminent characteristics:
the existence of a natural cutoff in the order of the Planck length $l_{Pl}$ which represents
the minimal accessible distance (see, e.g., \cite{GUP0}-\cite{GUP12}) and deviation from the standard relativistic
dispersion relation. This implies that at regions
dominated by QG effects, the relativistic dispersion relation should be modified. One of these approaches
in which both the attributes noted above are addressed was presented by Amelino-Camelia \cite{Ame0, Ame1} and is knows
as ``Doubly Special Relativity'' (DSR). Specifically, DSR is an advanced version of special relativity
in the presence of an excess invariant object named Planck energy $E_{Pl}$ \cite{Mag0, Mag1, Mag2}. Following this
proposal, Magueijo and Smolin \cite{Mag3} generalized DSR to  ``Doubly General Relativity'' (DGR) through involvement of gravity. The core idea of the DGR is that at high energy regimes we will not meet a single geometry of classical space-time, rather it is a ``running geometry''. Indeed, the geometry of space-time is detected by the energy dependent quantum particle(s) known as ``probe particle(s)'' which are traveling in it. As in quantum mechanics (QM) where the system under measurement has interaction with measuring device, the classical background geometry can be affected by the movements of these probe particle(s) with various energies and interactions between them which would lead to changes in the standard relativistic dispersion relation, written as
\begin{equation}\label{e1-1}
E^{2}f_{1}\left(\frac{E}{E_{Pl}}\right)-p^{2}f_{2}\left(\frac{E}{E_{Pl}}\right)=m^2~.
\end{equation}
Depending on the energy level with which  the space-time is explored,
i.e. the value of dimensionless ratio $\frac{E}{E_{Pl}}$, probe particles
record the various pictures of the space-time background. Inspired by such a
feature, the DGR scenario is known as the ``gravity's rainbow.''
Accordingly, $f_{1,2}\left(\frac{E}{E_{Pl}}\right)$ in the modified dispersion
relation (MDR) (\ref{e1-1}), are called the ``rainbow functions'' and have a two-facet
significance;  they lead to a MDR which in one hand must be consistent with some
outcomes of other QG approaches and on the other hand they should help in resolving the paradoxes
created in justifying some of the cosmological phenomenon. It should be noted that to respect
the usual formula, the rainbow functions should obey
\begin{equation}\label{e1-2}
\lim_{\frac{E}{E_{Pl}}\,\rightarrow\,0}f_{1,2}\left(\frac{E}{E_{Pl}}\right)=1~.
\end{equation}

In this work, we focus attention on one of the cosmological applications of this semi-classical QG
approach, namely  the status of the initial singularity problem by taking the quantum
corrections allowed by the DGR proposal. Numerous works have been carried out in recent years regarding
gravity's rainbow proposal. For instance in \cite{Ali},  the authors study a rainbow FRW cosmology which
is fixed by two rainbow functions and derive some non-singular analytical solutions. In \cite{Bar}, a quantum cosmological perfect fluid model is considered in the context of rainbow gravity and the possibility of avoiding the initial singularity is studied, leading to a solution predicting the existence of a bouncing non-singular universe. In addition to the mechanism of establishing the potential barrier to remove the initial singularity, there is another idea known as the static universe (SU) which is extensively discussed in recent years. Based on this scenario, the present universe could have been commenced from an initial frozen state known as the SU at the asymptotic earliest times \cite{Vilen0, Vilen1} in the absence of the Big Bang. In other words, the heart of this scenario is that there is no time origin for the beginning of the universe. In another scenario known as the emergent cosmology (EC) \cite{Talk0, Talk1, Talk2}, one considers a closed universe having a positive curvature constant with a primary origin from which the SU begins where there are no issues such as initial singularity and horizon \cite{Ellis1, Ellis2}. Note that, while the positive spatial curvature idea has no consequence at late time cosmology, it can address some fundamental problems of GR in the early universe. Of course, this idea is endorsed implicitly by observational data since observations have revealed that we do not live in an exactly
flat universe  \cite{CMB0, CMB1}. The classical stability of the initial SU with respect to perturbations of the scalar and tensor modes within the framework of the DGR proposal have been studied by fixing two distinct rainbow functions \cite{Mohsen1}. Depending on the type of rainbow function, one meets the different stability conditions. It would
therefore be of interest to investigate the \emph{quantum stability} of rainbow cosmology.  More precisely, in this paper we endeavor to find an answer to the question of whether or not the closed rainbow cosmology remains stable from the QM viewpoint. To this end, we have limited our analysis to a minisuperspace model containing one degree of freedom, i.e. scale factor of the universe.

The plan of the paper is as follows: in Section 2 we derive the classical Hamiltonian by means of the Schutz's formalism for an ideal fluid plus cosmological constant. Section 3 deals with a closed rainbow universe model with two components of matter sources; dust and the cosmological constant. By analyzing the potential part of classical Hamiltonian and  WKB approximation, we examine the quantum stability of non-singular solutions. We then move on to consider a closed rainbow FRW universe model including an exotic matter field as a domain wall fluid plus the  cosmological constant. Our analysis on the initial singularity of the model in the context of gravity's rainbow cosmology finishes in section 4 by the
quantization of the classical Hamiltonian and consequently derivation of the wave function of the
universe. Section 5 is devoted to conclusions.\\

\section{The Hamiltonian}
We start by considering the action rose up from ADM formalism for gravity in the
presence of a cosmological constant $\Lambda$ and perfect fluid as
\begin{equation}\label{e2-1}
S = \int_M d^4 x \sqrt{-g}~ R + 2 \int_{\partial M} d^3 x \sqrt{h}~ h_{ab}~K^{ab} + \int_M d^4
x\sqrt{-g}~(P+\Lambda)~,~~ (16\pi G=\hbar=1)~.
\end{equation}
Here $h_{ab}$ represents the induced metric over three dimensional spatial hypersurface,
$K^{ab}$ is the extrinsic curvature and $P$ is the pressure defined via the usual equation
of state (EOS) $P=\omega\rho$ with $\rho$ being the energy density. So one see in the above action included the
perfect fluid energy density inside the Lagrangian. In \cite{PH} has been presented
an action known as Hawking-Ellis as $S_M=-\int d^4x\sqrt{-g}\zeta(1+\upsilon)$ corresponding
to the action of perfect fluid so that in which $\zeta$ and $\upsilon$ represent the
density of fluid's and internal energy, respectively. In the same reference
it is proven that in the case of defining $\rho=\zeta(1+\upsilon)$ and $P=\zeta^2\frac{d
\upsilon}{d\zeta}$, by varying the mentioned action in terms of metric obtain the same standard
expression for the energy-momentum tensor. Surprisingly, in \cite{sch0, sch1} shown that
this result also can be derivable in the same way by considering the perfect fluid energy density
inside the Lagrangian. Note that the above action suggested according to ADM formalism so that
the seconded term - as a boundary term- will be remove via the variation of $\int_{M}d^{4}x
\sqrt{-g}R$, see \cite{Mohsen2} for more details.  For an isotropic and homogeneous universe the general form of the rainbow FRW metric reads
\begin{equation}\label{e2-2}
ds^2 = \frac{N^2(t)}{f_1^2(\frac{E}{E_{Pl}})} dt^2 - \frac{a^2(t)}{f_2^2(\frac{E}{E_{Pl}})}
\left[ \frac{dr^2}{1-kr^2}+ r^2 d\vartheta^2 + r^2 \sin^2 \vartheta~d\varphi^2 \right]~.
\end{equation}
Here $N$ represents the lapse function and $k$ takes one of the three values $-1,0,+1$ corresponding
to an open, flat or closed universe respectively. In the modified FRW metric (\ref{e2-2}), the quantum
corrections are embedded in the rainbow functions $f_{1,2}\left(\frac{E}{E_{Pl}}\right)$. In other words, these functions indicate how the standard FRW metric can be deformed as a consequence of the motion of quantum probe particles in the early classical space-time geometry. One my urge that $f_{1,2}\left(\frac{E}{E_{Pl}}\right)$ can be absorbed in the lapse function and scale factor respectively just by redefinition of these quantities and therefore there is no trace of energy-dependence in this metric in essence. While this seems to be the case for $f_{1}$, we note however that energy dependence of the background metric in gravity's rainbow needs a reconsideration of the measurement process and therefore one can not say that this is just a re-parametrization of the lapse function and scale factor. In order to calculate the Hamiltonian for the action (\ref{e2-1}) we start from the fluid part. Although the fluid's four velocity $U_{\nu}$ in Schutz's formalism \cite{sch0, sch1} is defined in terms of six potentials, it can be rewritten in terms of  four independent potentials
$h, S, \epsilon$ and $\theta$ as follows\footnote{Among the six potentials, two are linked to rotation.
The FRW type models, on the other hand, admit time-like geodesics which are hyper-surface normal i.e.
the vorticity tensor $\omega_{\mu\nu}$ is zero. Note that $\omega_{\mu\nu}$ refers to the rotation
of time-like geodesics.}
\begin{equation}\label{e2-3}
U_{\mu} = \frac{1}{h} (\epsilon_{,\mu} + \theta S_{,\mu}) ~.
\end{equation}
The fluid's four velocity $U_{\mu}$ obeys  $U_{\mu}U^{\mu} =1$. To make contact with thermodynamic quantities,
we can interpret $h$ and $S$ as the specific enthalpy and specific entropy, respectively. Therefore, the fluid
part of the action (\ref{e2-1}) can be rewritten using the following relevant thermodynamic equations \cite{sch}
\begin{eqnarray}\label{e2-4}
\begin{array}{ll}
\rho = \rho_0(1+ \Pi)~,\\
h = 1 + \Pi + \frac{P}{\rho_0}~,\\
\tau dS = d\Pi + P d \left(\frac{1}{\rho_0}\right)~.
\end{array}
\end{eqnarray}
where $\tau, \rho, \rho_0$ and $\Pi$ denote temperature, total mass energy density, rest mass density and specific
internal energy, respectively. By combining the thermodynamic relations given in (\ref{e2-4}) one can easily prove
that the EoS reads as
\begin{equation}\label{e2-5}
P= \frac{\omega}{(1+\omega)^{\frac{\omega+1}{\omega}}}h^{\frac{\omega+1}{\omega}}e^{-\frac{S}
{\omega}}~.
\end{equation}
In a comoving system with a perfect fluid four vector velocity $U_{\mu} = \left(N{f_1^{-1}(\frac{E}{E_{Pl}})},0,0,0\right)$
on can deduce the following Lagrangian
\begin{equation}\label{e2-6}
L_{f-\Lambda} = \frac{f_1^{\frac{1}{\omega}}(\frac{E}{E_{Pl}})}{f_2^3(\frac{E}{E_{Pl}})} ~a^3 N^{-\frac{1}
{\omega}}~\frac{\omega (\dot{\epsilon}+\theta \dot{S})^{\frac{\omega+1}{\omega}}}{(1+\omega)^{\frac{\omega+1}
{\omega}}}~ e^{-\frac{S}{\omega}} + \frac{N a^{3}}{f_1(\frac{E}{E_{Pl}})f_2^3(\frac{E}{E_{Pl}})}\Lambda ~.
\end{equation}
so that $h>0$ and $(\dot{\epsilon}+\theta \dot{S})>0$. Also, the Hamiltonian  takes the following form
\begin{equation}\label{e2-7}
H_{f-\Lambda} =  \dot{\epsilon}P_{\epsilon}+ \dot{S} P_{S} - L_{f-\Lambda}=
\frac{N f_2^{3\omega}(\frac{E}{E_{Pl}})}{f_1(\frac{E}{E_{Pl}})} \frac{P_T}{a^{3\omega}}
- \frac{N a^{3}}{f_1(\frac{E}{E_{Pl}})f_2^{3}(\frac{E}{E_{Pl}})}\Lambda~.
\end{equation}
As is clearly seen, the Hamiltonian is a linear function of $P_T$ such that $P_{\epsilon}=
\frac{\partial}{\partial \dot{\epsilon}}L_{f-\Lambda}$, $P_{S}=\frac{\partial}{\partial \dot{S}}L_{f-\Lambda}.$
To obtain the Hamiltonian in terms of $P_T$, we have employed the following canonical
transformations introduced in \cite{Lap}\footnote{Let us recall that by way of such canonical transformations, we
may pursue  the status of a dynamical system with more variables \cite{Babak, Mohsen2}.}
\begin{eqnarray}\label{e2-8}
\begin{array}{ll}
T=-P_S e^{-S}P_{\epsilon}^{-(\omega+1)}~,\\
P_T = P_{\epsilon}^{\omega+1} e^S~.
\end{array}
\end{eqnarray}
The Lagrangian and consequently the Hamiltonian corresponding to the gravity part of the
action (\ref{e2-1}) can be written as
\begin{equation}\label{e2-9}
L_g = -6 ~\frac{f_1(\frac{E}{E_{Pl}})}{f_2^3(\frac{E}{E_{Pl}})} ~\frac{a\dot{a}^2}{N} +
6 ~\frac{N~k~a}{f_1(\frac{E}{E_{Pl}})f_2(\frac{E}{E_{Pl}})}~,
\end{equation}
and
\begin{equation}\label{e2-10}
H_g = \dot{a} P_{a} -L_g = - \frac{Nf_2^3(\frac{E}{E_{Pl}})}{24f_1(\frac{E}{E_{Pl}})}
~\frac{P_a^2}{a} - \frac{6~N~k~a}{f_1(\frac{E}{E_{Pl}})f_2(\frac{E}{E_{Pl}})}~,
\end{equation}
respectively with $p_a=\frac{\partial L_g}{\partial \dot{a}}$ which is the momentum canonically
conjugated to the scale factor $a$. The super Hamiltonian for the minisuperspace of this model
can now be written as
\begin{equation}\label{e2-11}
H=H_g + H_{f-\Lambda} = \frac{N}{f_1(\frac{E}{E_{Pl}})} \left(-\frac{f_2^3(\frac{E}{E_{Pl}})}{24}
\frac{P_a^2}{a} -\frac{6ka}{f_2(\frac{E}{E_{P}})} + f_2^{3\omega}(\frac{E}{E_{Pl}})~ \frac{P_T}
{a^{3\omega}} - \frac{\Lambda a^{3}}
{f_2^{3}(\frac{E}{E_{Pl}})} \right) ~.
\end{equation}
In Eq. (\ref{e2-11}), $N$ is called the Lagrange multiplier preserving the classical constraint
equation $H=0$. Given that $\dot{T}=\{T,H\} = \frac{Nf_2^{3\omega}(\frac{E}{E_{Pl}})}{f_1(\frac{E}
{E_{Pl}})a^{3\omega}}$, $T$ in Eq. (\ref{e2-11}) might serve the role of the time (i.e. $T=t$)
provided that
\begin{equation}\label{e2-12}
N=\frac{f_1(\frac{E}{E_{Pl}})}
{f_2^{3\omega}(\frac{E}{E_{Pl}})}a^{3\omega}~.
\end{equation}
Thereupon, we can write the final form of the super Hamiltonian (\ref{e2-11}) as
\begin{equation}\label{e2-13}
H= f_{2}^{3-3\omega}\left(\frac{E}{E_{Pl}}\right)a^{3\omega-1} \left[\frac{p_a^2}{24} +
6ka^{2}{f_{2}^{-4}\left(\frac{E}{E_{Pl}}\right)} - a^{1-3\omega}f_{2}^{3\omega-3}\left(\frac{E}
{E_{Pl}}\right)p_T + a^{4}f_{2}^{-6}\left(\frac{E}{E_{Pl}}\right)\Lambda\right] ~.
\end{equation}

At this point an important issue should be noted. As the above relation indicates, the rainbow function $f_1(\frac{E}{E_{Pl}})$
does not contribute to the final form of the super Hamiltonian in the same way as the lapse function $N$ does. More precisely, since the lapse function is arbitrary,
the rainbow function $f_1(\frac{E}{E_{Pl}})$ can always be absorbed by re-scaling $N$. In fact, based on some symmetry properties of the spacetime in cosmological scales,
one can define the well known standard (or global) time coordinate, the  \emph{cosmic time} $\eta\equiv\int N dt$. Here, by rescaling the lapse function,
definition of the cosmic time is modified as $\eta=\int \frac{N}{f_1(\frac{E}{E_{Pl}})} dt$ which, based on the correspondence
principle, recovers the standard definition of the cosmic time in low energy scales. Using $N$ in (\ref{e2-12}),
one finds that the energy dependent function $f_1(\frac{E}{E_{Pl}})$ disappears and has no affect on the physical properties
or measurements expressed by the cosmic time coordinate $\eta$. In effect, it is equivalent to fixing the value of $f_1(\frac{E}{E_{Pl}})$
to unity in temporal part of the line element. One may argue that the results included in the present analysis, given
by two energy-dependent functions $f_1$ and $f_{2}$ in the line element (4), appear
to be a coordinate effect. While it is indeed the case for $f_{1}$ due to definition of time as we have explained, the situation is different for $f_{2}$.
In gravity's rainbow proposal the unknown energy-dependent function $f_2(\frac{E}{E_{Pl}})$ has physical implications and changes measurement process in essence.
The physical roles of this function cannot be ignored by a simple rescaling of the scale factor. For instance, the scale factor rescaled by
$f_2(\frac{E}{E_{Pl}})$ in the spatial part of the line element (4) has the potential to remove the initial singularity
in the cosmic history, see \cite{Li} for details. Indeed, Magueijo and Smolin in their seminal work \cite{Mag2}, by using rescaling of the time coordinate  $\tau(E)=\frac{h(E)}{f(E)}t$ via introducing some unknown energy dependent functions $h$ and $f$, have opened novel avenues for the solution of the horizon problem. It is noteworthy that here the solution of the horizon problem is subject to an appropriate choice of the above mentioned rainbow functions.  In summary, the role of the rainbow function $f_2$ cannot be reduced to
merely a coordinate transformation and has significant effects on the cosmological scenario under consideration. In what follows,  one of the its cosmological application will be discussed in details.

\section{Quantum stability of closed rainbow cosmology}
Based on classical theory, rainbow universe can be considered as
a constrained dynamical system resulting from Hamiltonian
(\ref{e2-13}). The constraint equation $H=0$ allows us to rewrite
 Hamiltonian (\ref{e2-13}) in terms of the kinetic and potential
energies as follows
\begin{equation}\label{e3-1}
p_{a}^{2}+V\left(a,\frac{E}{E_{Pl}}\right)=0~,
\end{equation}
so that
\begin{equation}\label{e3-2}
p_{a}= -12~a^{1-3\omega}~\dot{a}~f_2^{3\omega-3}\left(\frac{E}{E_{Pl}}\right)~,
\end{equation}
and
\begin{equation}\label{e3-3}
V\left(a,\frac{E}{E_{Pl}}\right)= 144k{f_{2}^{-4}\left(\frac{E}{E_{Pl}}\right)}a^{2} -
24f_{2}^{3\omega-3}\left(\frac{E}{E_{Pl}}\right)p_T a^{1-3\omega} + 24f_{2}^{-6}
\left(\frac{E}{E_{Pl}}\right)\Lambda a^{4}~.
\end{equation}
The benefit of the above decomposition is that we can imagine the universe
as a non-relativistic particle which is under the influence of the one-dimensional
potential (\ref{e3-3}). We note that from a classical  viewpoint, regions $V\leq0$
are accessible to the traveling particle, here the universe. In what follows, by taking
two common rainbow functions  for two universe models filled with matter
sources mentioned in Section 2, we survey the stability of a closed rainbow
universe from  a QM viewpoint.

\subsection{Non-relativistic dust matter plus cosmological constant $\Lambda$}
For a closed rainbow universe model with a source of attraction as non-relativistic dust,
i.e. $\omega=0$, along with cosmological constant\footnote{While the sign of the cosmological
constant $\Lambda$ is not yet clear, an analysis of the potential part of the Hamiltonian in the
presence of the two known rainbow functions leads to its determination.} (attraction or repulsion),
the potential function (\ref{e3-3}) becomes
\begin{equation}\label{e3-4}
V\left(a,\frac{E}{E_{Pl}}\right)= 24f_{2}^{-6}\left(\frac{E}{E_{Pl}}\right) \Lambda a^{4}
+144{f_{2}^{-4}\left(\frac{E}{E_{Pl}}\right)}a^{2} - 24f_{2}^{-3}\left(\frac{E}{E_{Pl}}\right)p_T ~a ~.
\end{equation}
As previously mentioned, finding a relevant and useful rainbow function from both
theoretical and phenomenological considerations is one of the open issues in rainbow
gravity proposal. Here, we restrict ourselves to two most widely used rainbow functions
\begin{equation}\label{e3-5}
f_{2}\left(\frac{E}{E_{Pl}}\right)=\left(1-\frac{E}{E_{Pl}}\right)^{-1}~,
\end{equation}
and
\begin{equation}\label{e3-6}
f_{2}\left(\frac{E}{E_{Pl}}\right)=\sqrt{1-\left(\frac{E}{E_{Pl}}\right)^n}~,~~~~~~n=1,2,3, ...
\end{equation}
which have been suggested in \cite{Mag1} and \cite{Amelino1}, respectively.
Let us emphasize that either choice of the above functions does not mean that
they are problem free. However, their phenomenological aspects have particular
importance. For instance, the MDR obtained using the rainbow function (\ref{e3-6})
with $f_{1}(\frac{E}{E_{Pl}})=1$ to describe many of the phenomenon of interest
seems to work well \cite{PI}.  Now, starting from the rainbow function (\ref{e3-5}),
we arrive at
\begin{equation}\label{e3-7a}
V\left(a,\frac{E}{E_{Pl}}\right)= 24\left(1-\frac{6E}{E_{Pl}}\right) \Lambda a^{4}
+144\left(1-\frac{4E}{E_{Pl}}\right)a^{2} - 24\left(1-\frac{3E}{E_{Pl}}\right)p_T
~a ~.
\end{equation}
Through the minimization of the above potential function we find that
under the conditions listed in Eq. (\ref{e3-7b}), we are dealing with an
early static universe (SU) ($V_{min}=0$ ) which is fluctuating in the vicinity of non-zero
values of the scale factor $a_{0}$
\begin{eqnarray}\label{e3-7b}
\left\{
\begin{array}{ll}
p_{T}=0,~~ \frac{1}{6}<\frac{E}{E_{Pl}}\leq\frac{1}{4},~~ \Lambda \leq0, \\\\
p_{T}=0,~~ 0\leq\frac{E}{E_{Pl}}<\frac{1}{6},~~ \Lambda\geq0~,  \\\\
p_{T}>0,~~  \frac{1}{3}<\frac{E}{E_{Pl}}\leq1,~~ \Lambda\leq F(\frac{E}{E_{Pl}})~, \\\\
p_{T}\leq0,~~ 0\leq\frac{E}{E_{Pl}}<\frac{1}{6},~~ \Lambda\geq0~, \\\\
p_{T}<0,~~ \frac{1}{6}<\frac{E}{E_{Pl}}\leq\frac{1}{4},~~ \Lambda\leq0~.  \\\\
\end{array}
\right.
\end{eqnarray}
where $F(\frac{E}{E_Pl})$ reads as
\begin{equation}\label{}
F(\frac{E}{E_{Pl}})=\frac{-2048(\frac{E}{E_{Pl}})^3
+1536(\frac{E}{E_{Pl}})^2-384(\frac{E}{E_{Pl}})+32}{\left(54(\frac{E}{E_{Pl}})^3-45
(\frac{E}{E_{Pl}})^2+12(\frac{E}{E_{Pl}})-1\right)p_{T}^2}~.
\end{equation}
We see from Figure 1 (left panel) that around $a_{0}\sim2-2.5$, a SU takes shape.
While from the classical viewpoint, $a_{0}$ is a half stable point, this figure
reflects the fact that from  the QM viewpoint, there is the chance of a tunneling
via the potential barrier to a zero value of the scale factor. Surprisingly, Figure 1
(left panel) shows qualitatively that as the dimensionless ratio $\frac{E}{E_{Pl}}$ grows,
the barrier height also increases and the chance of tunneling becomes tiny. We
know from ordinary QM that the probability of tunneling is given in the
WKB-approximation by \cite{Dab, Atk}
\begin{equation}\label{e3-8}
\emph{{\cal \textbf{P}}}\sim~e^{-2\cal{I}}~,
\end{equation}
where $\cal{I}$ is called WKB tunneling action and is defined as
\begin{equation}\label{e3-9}
{\cal{I}}=\int_{0}^{a_{0}}\sqrt{V}~da~.
\end{equation}
The third constraints in Eq. (\ref{e3-7b}) is very suitable in the sense that, unlike
other circumstances, it allows us to explore the geometry of space-time by high energy
particle probes. Substituting the potential function (\ref{e3-7a}) in (\ref{e3-9}) and
using parameter values $p_{T}=15$, $a_{0}\sim 2$ with $\Lambda=F(\frac{E}{E_Pl})$ allows
us to perform a numerical analysis based on Eq. (\ref{e3-8}), see Table 1. Indeed, this
analysis represents the possibility of the transition of SU through the barrier to a zero
value of the scale factor.
\begin{table}
\caption{\footnotesize Numerical analysis of the probability of the SU collapse using
WKB-approximation for potential $V$, Eq. (\ref{e3-7a}).  To satisfy the third
constraint in (\ref{e3-7b}), use has been made of $\Lambda=F(\frac{E}{E_{Pl}})$ and
the parameter value $p_{T}=15$.}
\begin{center}
\begin{tabular}{|c|c|c|c|}
  \hline
   1 & $\frac{E}{E_{Pl}}$& $\cal{I}$ & $\emph{{\cal \textbf{P}}}$	 \\ \hline
   2 & $0.4$& $31.5648$ & $e^{-63.1296}$	 \\ \hline
   3 & $0.45$& $36.1046$ & $e^{-72.2092}$	 \\ \hline
   4 & $0.5$& $63.3813$ & $e^{-126.763}$	 \\ \hline
   5 & $0.55$& $96.5461$ & $e^{-193.092}$	 \\ \hline
   6 & $0.6$& $132.062$ & $e^{-264.124}$	 \\ \hline
   7 & $0.65$& $168.745$ & $e^{-337.49}$	 \\ \hline
   8 & $0.7$& $206.091$ & $e^{-412.182}$	 \\ \hline
   9 & $0.75$& $243.848$ & $e^{-487.696}$	 \\ \hline
   10 & $0.8$& $281.878$ & $e^{-563.756}$	 \\ \hline
        \end{tabular}
  \end{center}
\end{table}
In agreement with Figure 1 (left panel), values reported in Table 1 suggest that from QM
viewpoint, increasing the energy levels of the probe particle(s) makes the possibility
of the SU collapse via quantum tunneling small. Overall, it can be said that for a
rainbow universe model filled with dust plus cosmological constant, by satisfying
the third constraint in (\ref{e3-7b}), we are dealing with an early SU in the
vicinity of a non-zero scale factor $a_{0}\neq0$ which avoids the singularity.
The same results obtained from Figure 1 (left panel) and Table 1, can be summarized
as follows: the point $a_{0}\neq0$ may be stable quantum mechanically. What
should be noted is that each of the numerical values presented in the above Table
and also next Tables are not particularly illuminating. In fact, their up or down trends
in terms of dimensionless ratio $\frac{E}{E_{Pl}}$ point to a physical
interpretation. Recall
that the SU fluctuating in the vicinity of the point $a_{0}$
does not have exact classical stability, rather it is a half stable point.

The conditions below imply that for the potential function (\ref{e3-7a}) there is
a non zero value in the classically allowed region $V<0$.
\begin{eqnarray}\label{e3-7c}
\left\{
\begin{array}{ll}
p_{T}=0,~~ \frac{1}{4}<\frac{E}{E_{Pl}}\leq1,~~ \Lambda <0~, \\\\
p_{T}>0,~~ \frac{1}{6}<\frac{E}{E_{Pl}}<\frac{1}{4},~~ \Lambda =0~,   \\\\
p_{T}>0,~~ \frac{E}{E_{Pl}}=\frac{1}{3},~~ \Lambda <0~. \\\\
\end{array}
\right.
\end{eqnarray}
The first constraint given in Eq. (\ref{e3-7c}) is the most suitable constraint for
exploring the geometry of classical space-time background at energies comparable to
Planck energy. It suggests a harmonic universe in the presence of an initial singularity
which is oscillating between two classical turning points $a_1$ and $a_{2}$, so that
\begin{equation}\label{e3-10}
a_1=0,~~~~a_{2}=\sqrt{\frac{6\left(\frac{4E}{E_{Pl}}-1\right)}{\Lambda\left(1-\frac{6E}{E_{Pl}}
\right)}}~.
\end{equation}
We see from relation obtained above and as well as Figure 1 (right panel) that by increasing the
dimensionless ratio $\frac{E}{E_{Pl}}$ the point $a_{1}$ also shifts towards greater values.
Now, let us determine the status of the classical stability of the points $a_1=0$ and $a_{2}$.
The classical turning point $a_1=0$ is an unstable saddle point since $\frac{d}{da}V\left(a,\frac{E}
{E_{Pl}}\right)|_{a=a_1}=0=\frac{d^{2}}{da^{2}}V\left(a,\frac{E}{E_{Pl}}\right)|_{a=a_1}$. The
point $a_2$ also is an unstable point since $\frac{d}{da}V\left(a,\frac{E}{E_{Pl}}\right)|_{a=a_2}
<0$.

Another set of constraints derived from the analysis of the potential function (\ref{e3-7a})
together with the rainbow function  (\ref{e3-5}) can be written as
\begin{eqnarray}\label{e3-7d}
\left\{
\begin{array}{ll}
p_{T}=0,~~\frac{1}{4}<\frac{E}{E_{Pl}}\leq1,~~ \Lambda \geq0~,	 \\\\
p_{T}\geq0,~~ \frac{1}{6}<\frac{E}{E_{Pl}}<\frac{1}{4},~~ \Lambda>0~,  \\\\
p_{T}>0,~~ \frac{1}{3}\leq\frac{E}{E_{Pl}}\leq1,~~ \Lambda\geq0~, \\\\
p_{T}<0,~~ \frac{1}{6}<\frac{E}{E_{Pl}}\leq\frac{1}{4},~~ \Lambda>0~. \\\\
\end{array}
\right.
\end{eqnarray}
Constraints presented in (\ref{e3-7d}) indicate that there is no minimum for the potential
function (\ref{e3-7a}). We also note that the potential is commenced from a zero value for
the scale factor and  grows eventually to $|V|=\infty$. As mentioned earlier, in the context
of the semi-classical QG approaches, it is believed that the introduction of a non-zero minimal
length would render the initial singularity avoidable. Therefore, under constraints given in Eqs.
(\ref{e3-7c}) and (\ref{e3-7d}), the rainbow cosmology filled with matter sources considered above
is not devoid of the initial singularity.
\begin{figure}
\begin{tabular}{c}\hspace{-1cm}\epsfig{figure=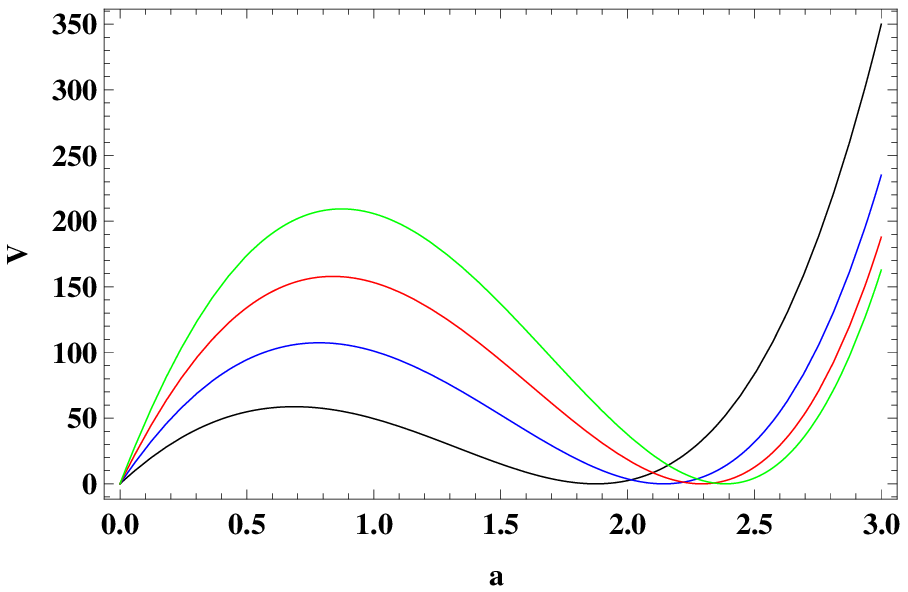,width=7cm}
\hspace{1cm} \epsfig{figure=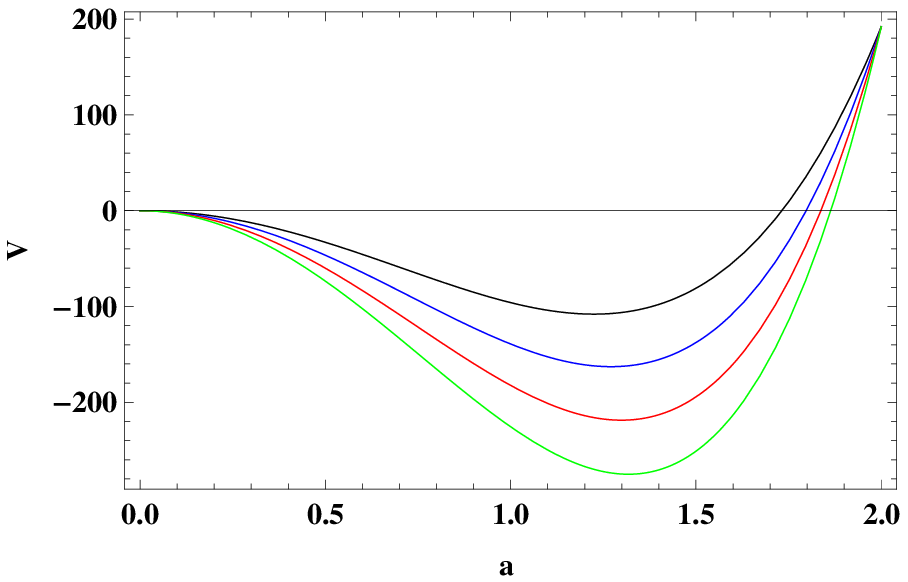,width=7cm}
\end{tabular}
\caption{\footnotesize  \textbf{Left}: Potential $V$, Eq. (\ref{e3-7a}) as a function of
scale factor $a$ for various values of dimensionless ratio $\frac{E}
{E_{Pl}}= 0.5$,\, 0.6$,\, 0.7, \,0.8$,
from bottom to top respectively. To satisfy the third constraint in (\ref{e3-7b}), we have used the
parameter value  $P_T= 15$. \textbf{Right}: Various values of dimensionless ratio $\frac{E}
{E_{Pl}}=  0.5,\, 0.6,\, 0.7, \, 0.8$, from top to bottom to satisfy the first constraint in Eq.
(\ref{e3-7c}). We have used the parameter values $P_T= 0$ and $\Lambda=-1$.}
\label{fig1}
\end{figure}
Let us now pursue our analysis using the rainbow function (\ref{e3-6}). This time the potential
function (\ref{e3-4}) can be rewritten as
\begin{equation}\label{e3-12}
V\left(a,\frac{E}{E_{Pl}}\right)= 24\left(1+3(\frac{E}{E_{Pl}})^n\right) \Lambda a^{4}
+144\left(1+2(\frac{E}{E_{Pl}})^n\right)a^{2} - 24\left(1+\frac{3}{2}(\frac{E}{E_{Pl}})
^n\right)p_T ~a ~.
\end{equation}
By the same method, we go through minimizing the potential function (\ref{e3-12}) and examine the
initial singularity of the rainbow cosmology. While we know that the value of the dimensionless
ratio $\frac{E}{E_{Pl}}$ lies in the interval $0\leq\frac{E} {E_{Pl}}\leq1$, we find that for the
odd values of $n$ minimization of the potential (\ref{e3-12}) leads to the constraint $\frac{E}{E_{Pl}}<0$,
which is not allowed. Also, for the even values of $n$ under no circumstances will we encounter
a SU ($V_{min}=0$) in the early universe.  This is while for the following condition there is
a non-zero minimum value in the classically allowed region $V<0$
\begin{equation}\label{e3-13}
p_{T}>0~,~~~~~\Lambda =0~,
\end{equation}
so that $V_{min}<0$.  Akin to Figure 1 (right panel), this condition also leads to an early singular
harmonic universe which is oscillating between $a_1$ and $a_{2}$, so that
\begin{equation}\label{e3-16}
a_1=0,~~~~a_{2}=\frac{p_{T}}{6}~\frac{\left(1+\frac{3}{2}(\frac{E}{E_{Pl}})^n\right)}{\left(1+2(\frac{E}
{E_{Pl}})^n\right)},~~~n=2,4,...~.
\end{equation}
However, for the following constraint
\begin{equation}\label{e3-17}
p_{T}\leq0~,~~~~~\Lambda <0~,
\end{equation}
there is an early potential barrier for  non-zero values of the scale factor $a_{2}\neq0$ which creates
a repulsive force preventing the formation of the initial singularity, see Figure 2 (left and right panels). We
see from this figure that by increasing the dimensionless ratio $\frac{E}{E_{Pl}}$ (left panel) and also the
canonically conjugate momenta to $T$ (right panel), the chance of quantum mechanical penetration via the potential
barrier from $a_{2}\neq0$ to $a_1=0$, becomes small. Interestingly, in Figure 2 (right panel) we see that for a fixed
value of $\frac{E}{E_{Pl}}$, an increase in $|p_{T}|$ causes a repulsive force to arise from the potential barrier
which appears at a greater minimum scale factor. Therefore, unlike the rainbow function (\ref{e3-5}), here the
initial singularity gets eliminated through a pure repulsion mechanism and causes the formation of a potential
barrier for non-zero values of the scale factor. Of course, from a classical dynamics approach, it is easily
recognizable that the point $a_2$ is stable.
\begin{figure}
\begin{tabular}{c}\hspace{-1cm}\epsfig{figure=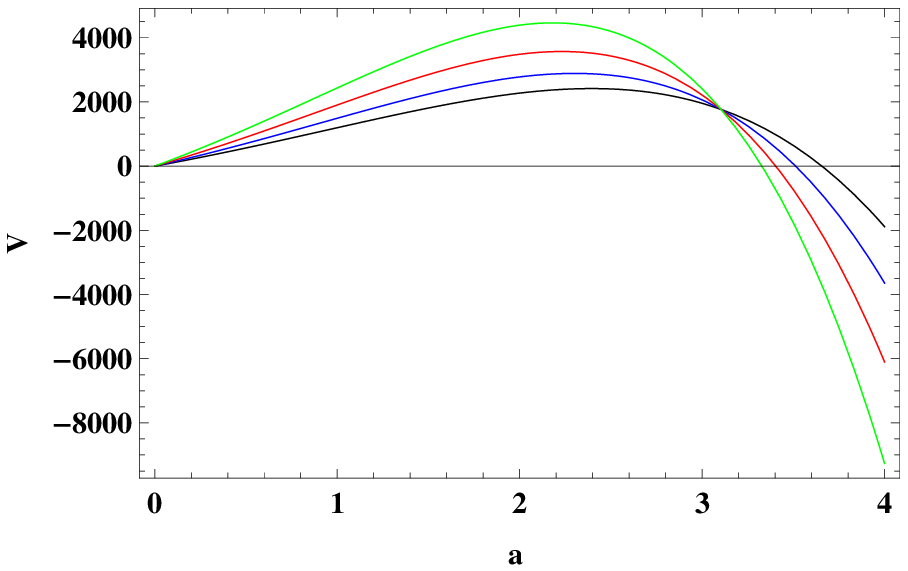,width=7cm}
\hspace{1cm} \epsfig{figure=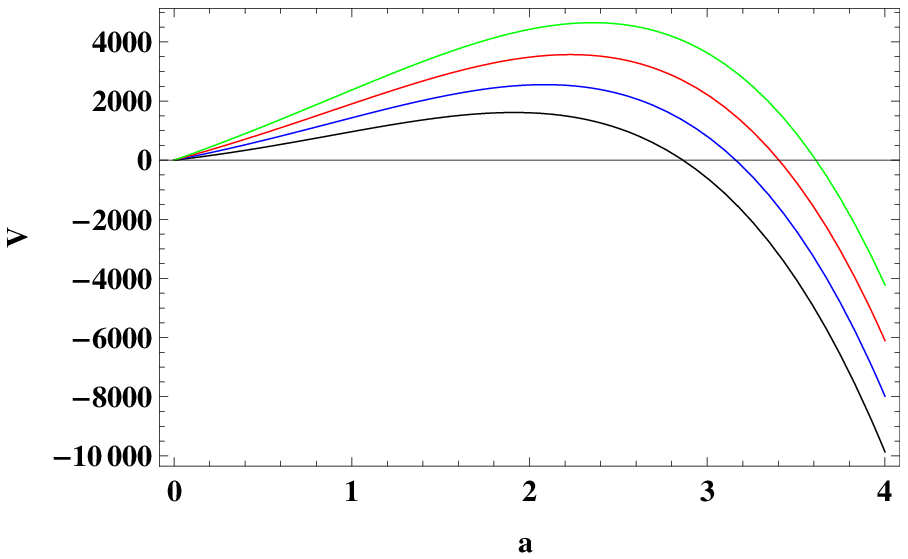,width=7cm}
\end{tabular}
\caption{\footnotesize  \textbf{Left}: The potential $V$, Eq. (\ref{e3-12}) as a function of the scale factor
$a$ for various values of dimensionless ratio $\frac{E}{E_{Pl}}=0.4,\, 0.6,\, 0.8, \,1$, from bottom to top
respectively. To satisfy constraint
Eq.(\ref{e3-17}) we have used parameters values $P_T= -35$, $\Lambda=-1$ and $n=2$. \textbf{Right}: Various
values of $P_T=-15,\, -25,\, -35, \,-45$ from bottom to top respectively.  We have used parameter
values $\frac{E}{E_{Pl}}=0.8$, $\Lambda=-1$ and $n=2$.}
\label{fig2}
\end{figure}

\subsection{Exotic matter as domain wall fluid plus cosmological constant $\Lambda$ }
Our goal here is to examine the initial singularity of rainbow cosmology through
analyzing the potential part of a one-dimensional minisuperspace model with a
dimensionless ratio $\frac{E}{E_{Pl}}$. This time however it is done in the context
of a closed rainbow FRW universe model filled with a repulsive (exotic) source as
a domain wall fluid with $\omega =-\frac{2}{3}$ \footnote{We know that a perfect fluid
with negative pressure results in instabilities on the very short scales. Nevertheless
if we assume the dark matter is a solid, with an elastic resistance to pure shear
deformations, then such short wavelength instabilities are avoidable by an EOS parameter
$\omega<0$ \cite{BUC}. One of possible candidates for a solid dark matter component
is a frustrated network of domain walls with fixed EOS parameter $\omega=-\frac{2}{3}$. In fact,
domain walls are topological defects which are expected to be formed along the phase
transitions in the initial moments of universe as it expands. Historically, the idea
was raised for the first time in $1974$ that a domain walls structure may be occur in
the framework of theories with spontaneous symmetry breaking \cite{1974}. Technically,
the phase transition happens due to losing the temperatures of universe below the threshold
associated with a certain Higgs field with non zero vacuum expectation value. Given that
the universe is so big, it is a reasonable expectation that in separated regions vacuum
expectation value not be the same. Therefore, these regions take arbitrarily different
expectation vacuum values along the phase transition so that the domain walls form at the
interface their between. It is interesting to note that present cosmological data highly
suggest a slight diversion of $\Lambda \mbox{CDM}$ models towards $\omega>-1$.}
plus a cosmological  constant $\Lambda$.
Therefore, Eq. (\ref{e3-3}) reads
\begin{equation}\label{e3-18}
V\left(a,\frac{E}{E_P}\right)= 24f_{2}^{-6}(\frac{E}{E_{P}})\Lambda a^{4} -
24f_{2}^{-5}(\frac{E}{E_{Pl}})p_T~ a^{3}+ 144{f_{2}^{-4}(\frac{E}{E_{P}})}a^{2} ~,
\end{equation}
where by substituting the rainbow functions (\ref{e3-5}) and (\ref{e3-6})
we arrive at
\begin{equation}\label{e3-19}
V\left(a,\frac{E}{E_{Pl}}\right)= 24(1-\frac{6E}{E_{Pl}}) \Lambda a^{4}
- 24(1-\frac{5E}{E_{Pl}})p_T ~a^3 + 144(1-\frac{4E}{E_{Pl}})a^{2}~,
\end{equation}
and
\begin{equation}\label{e3-19a}
V\left(a,\frac{E}{E_{Pl}}\right)= 24\left(1+3(\frac{E}{E_{Pl}})^n\right) \Lambda a^{4}
- 24\left(1+\frac{3}{2}(\frac{E}{E_{Pl}})^n\right) p_T ~a^3 ~+144\left(1+2(\frac{E}{E_
{Pl}})^n\right)a^{2}~.
\end{equation}
respectively.  As before, by means of minimizing the potential function (\ref{e3-19}),
we note that for the underlying constraints
\begin{eqnarray}\label{e3-20}
\left\{
\begin{array}{ll}
p_{T}=0,~~ \frac{1}{6}<\frac{E}{E_{Pl}}\leq\frac{1}{4},~~ \Lambda \leq0~, \\\\
p_{T}>0,~~ 0\leq\frac{E}{E_{Pl}}<\frac{1}{6}, ~~\Lambda\geq\frac{25(\frac{E}{E_{Pl}})^2
-10(\frac{E}{E_{Pl}})+1}{\left(576(\frac{E}{E_{Pl}})^2-240(\frac{E}{E_{Pl}})+24\right)
p_{T}^2}~.
\end{array}
\right.
\end{eqnarray}
there is a zero minimum value which points to the presence of an early SU fluctuating in
the vicinity of a non-zero value of the scale factor $a_{0}\neq0$. Due to the two constraints
listed above, the space-time background is just restricted to low energy levels. Now,
using the parameter value $P_T= 10$ with
\begin{equation}\label{e3-20b}
\Lambda
=\frac{25(\frac{E}{E_{Pl}})^2 -10(\frac{E}{E_{Pl}})+1}{\left(576(\frac{E}{E_{Pl}})
^2-240(\frac{E}{E_{Pl}})+24\right) p_{T}^2}~,
\end{equation}
to satisfy the second constraint, the shape of the potential function (\ref{e3-19}) is shown
in Figure 3 (left panel). It is easy to see that by following the rainbow gravity proposal to GR
in the limit $\frac{E}{E_{Pl}}\rightarrow0$, the chance of SU collapsing via quantum tunneling
becomes more pronounced. This result looks interesting in the sense that despite the classical
stability of a closed standard universe (including the exotic matter) and in the light of scalar
and tensor perturbations \cite{Ellis1, Ellis2}, it would not remain stable in the context of QM.
To verify this in a more accurate manner using WKB-approximation which was discussed earlier,
we present a numerical analysis of the probability of quantum tunneling (with the same fixed
numerical values as in Figure 3 (left panel)), in Table 2. To perform this task, the SU is fixed in
the vicinity of $a_{0}\sim2$. As can be seen, in accord with Figure 3, as dimensionless ratio
$\frac{E}{E_{Pl}}$ decreases the chance of SU collapsing grows.
\begin{table}
\caption{\footnotesize Numerical analysis of the probability of the SU collapse
using WKB-approximation for  potential $V$ (\ref{e3-19}). To satisfy
the second constraint in (\ref{e3-20}) we have used Eq.(\ref{e3-20b})
and the parameter value
$p_{T}=10$.}
\begin{center}
\begin{tabular}{|c|c|c|c|}
  \hline
   1 & $\frac{E}{E_{Pl}}$& $\cal{I}$ & $\emph{{\cal \textbf{P}}}$	 \\ \hline
   2 & $\frac{1}{7}$& $-109.702$ & $e^{219.404}$	 \\ \hline
   3 & $\frac{1}{8}$& $-167.982$ & $e^{335.964}$	 \\ \hline
   4 & $\frac{1}{9}$& $-213.311$ & $e^{426.621}$	 \\ \hline
   5 & $\frac{1}{10}$& $-249.573$ & $e^{499.147}$	 \\ \hline
   6 & $\frac{1}{11}$& $-279.243$ & $e^{558.486}$	 \\ \hline
   7 & $\frac{1}{12}$& $-303.967$ & $e^{607.935}$	 \\ \hline
   8 & $\frac{1}{13}$& $-324.888$ & $e^{649.776}$	 \\ \hline
   9 & $\frac{1}{14}$& $-342.82$ & $e^{685.64}$	      \\ \hline
   10 & $\frac{1}{15}$& $-358.361$ & $e^{716.722}$	  \\ \hline
     \end{tabular}
  \end{center}
\end{table}
In what follows, we see that in the classically allowed region $V<0$ for
the potential (\ref{e3-19}), there is a non zero minimal value $V_{min}<0$
provided that the following conditions are satisfied
\begin{eqnarray}\label{e3-20a}
\left\{
\begin{array}{ll}
p_{T}=0,~~ \frac{1}{4}<\frac{E}{E_{Pl}}\leq1,~~ \Lambda <0~, \\\\
p_{T}>0,~ \frac{1}{4}<\frac{E}{E_{Pl}}\leq1,~ \Lambda =0~.
\end{array}
\right.
\end{eqnarray}
The requirements mentioned above represent a singular harmonic universe with
general behavior similar to Figure 1 (right panel). It is oscillating between two
classical turning points $a_1$ and $a_{2}$
\begin{equation}\label{e3-21}
a_1=0,~~~~a_{2}=\sqrt{\frac{6\left(\frac{4E}{E_{Pl}}-1\right)}{\Lambda\left(1-\frac
{6E}{E_{Pl}}\right)}}~,
\end{equation}
and
\begin{equation}\label{e3-22}
a_{2}=\frac{6}{p_{T}}\frac{\left(1-\frac
{5E}{E_{Pl}}\right)}{\left(1-\frac
{4E}{E_{Pl}}\right)}~.
\end{equation}
We note that unlike previous rainbow FRW cosmology models (consisting of an attractive
source as dust plus a cosmological constant) with the same rainbow function, the initial
singularity here may be eliminated only for probing particle(s) with intermediate energy
levels, see  constraints (\ref{e3-20}).
\begin{figure}
\begin{tabular}{c}\hspace{-1cm}\epsfig{figure=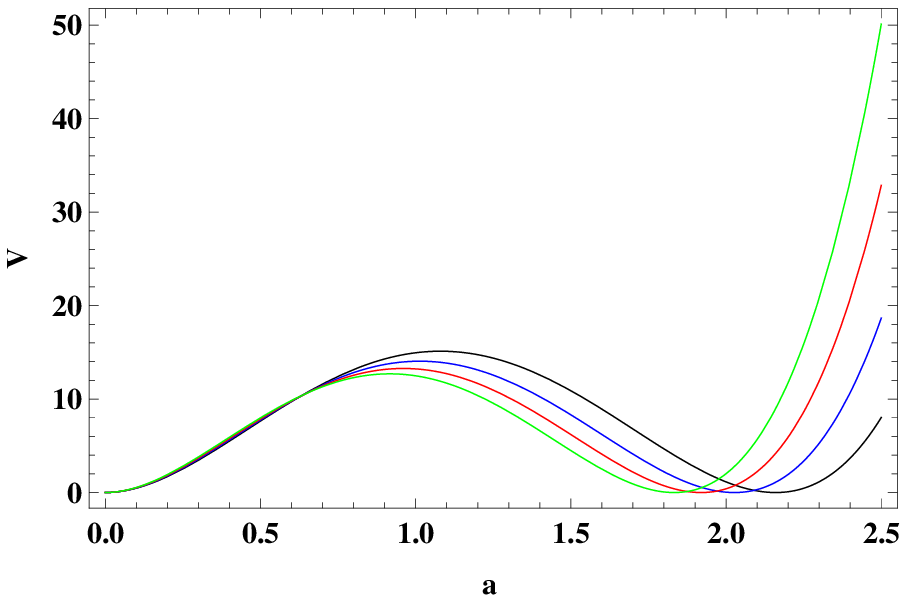,width=7cm}
\hspace{1cm} \epsfig{figure=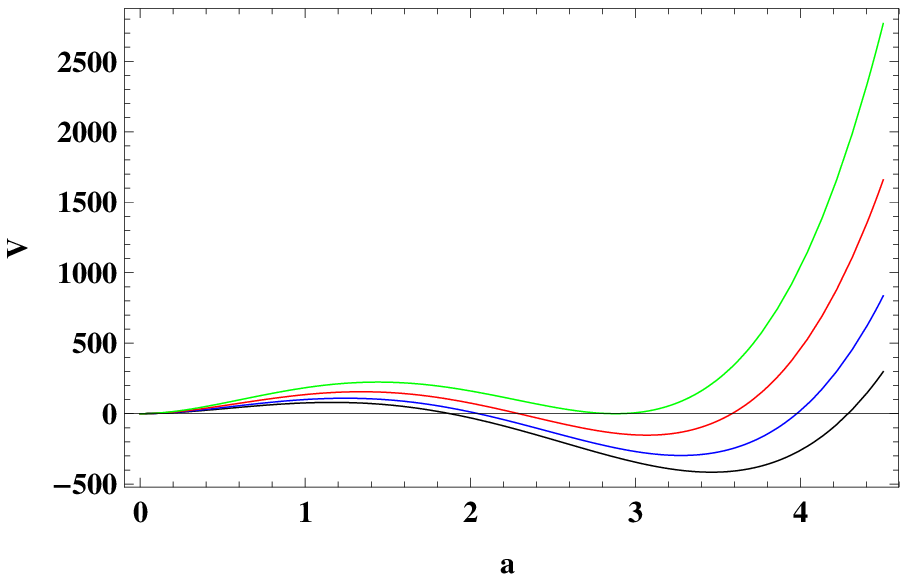,width=7cm}
\end{tabular}
\caption{\footnotesize  \textbf{Left}: the potential $V$, Eq. (\ref{e3-19}) as a function of scale
factor $a$ for various values of dimensionless ratio $\frac{E}{E_{Pl}}=0.160,\,0.155,\,0.150,\, 0.145$,
from top to bottom respectively. In plotting this figure, the satisfaction of the forth constraint in
(\ref{e3-20}) is taken into account. \textbf{Right}: The potential $V$, Eq. (\ref{e3-19a}) as a function of
scale factor $a$ for various values of dimensionless ratio $\frac{E}{E_{Pl}}=0.7,\,0.8,\,0.9,\,1$ from bottom
to top respectively. To satisfy constraint (\ref{e3-22}) we have used parameter values corresponding to $p_{T}=10$
and $n=4$.}
\label{fig3}
\end{figure}
\begin{table}
\caption{\footnotesize Numerical analysis of the probability of the SU collapse using WKB-approximation
for potential $V$ (\ref{e3-19a}). To satisfy  constraint (\ref{e3-22}), we have set
parameter values at $p_{T}=10$ and $n=4$.}
\begin{center}
\begin{tabular}{|c|c|c|c|}
  \hline
   1 & $\frac{E}{E_{Pl}}$& $\cal{I}$ & $\emph{{\cal \textbf{P}}}$	 \\ \hline
   2 & $0.6$& $63.4409$ & $e^{-126.882}$	 \\ \hline
   3 & $0.65$& $73.204$ & $e^{-146.408}$	 \\ \hline
   4 & $0.7$& $86.7097$ & $e^{-173.419}$	 \\ \hline
   5 & $0.75$& $105.072$ & $e^{-210.144}$	 \\ \hline
   6 & $0.8$& $129.74$ & $e^{-259.481}$	      \\ \hline
   7 & $0.85$& $162.598$ & $e^{-325.196}$	 \\ \hline
   8 & $0.9$& $206.146$ & $e^{-412.292}$	 \\ \hline
   9 & $0.95$& $264.009$ & $e^{-528.018}$	 \\ \hline
   10 & $1$& $343.985$ & $e^{-687.971}$	 \\ \hline
    \end{tabular}
  \end{center}
\end{table}
We first note that according to the analysis done on the potential function (\ref{e3-19a}),
for the same reason mentioned  previously, odd values of parameter $n$ are not allowed.
Secondly, under the following constraints
\begin{equation}\label{e3-22}
p_{T}>0,~~~~~~\Lambda=\frac{9(\frac{4E}{E_{Pl}})^4+12(\frac{4E}{E_{Pl}})^2+4}{576(\frac{4E}
{E_{Pl}})^4+480(\frac{4E}{E_{Pl}})^2+96}~p_{T}^{2}~,
\end{equation}
for the case\footnote{For the case $n=2$, constraints (\ref{e3-22}) result in having the
SU around a non-zero value of the scale factor akin to shapes displayed in Figures 1 and 3
(left panels).} $n=4$ and some values of $\frac{E}{E_{Pl}}$,  we are dealing with a harmonic
universe which oscillates between the minimum and maximum values of the scale factor $a_{1}$
and $a_{2}$, for which $V(a_{1,2})=0$
\begin{equation}\label{e3-23}
a_{1,2}=\frac{24 \left(18 (\frac{E}{E_{Pl}})^8 + 15 (\frac{E}{E_{Pl}})^6 + 15 (\frac{E}{E_{Pl}})^4
+ 10 (\frac{E}{E_{Pl}})^2 + 2  \mp A \right)}{\left(27(\frac{E}{E_{Pl}})^8 + 36 (\frac{E}{E_{Pl}})^6
+ 21 (\frac{4E}{E_{Pl}})^4 + 12 (\frac{4E}{E_{Pl}})^2 +2 \right)p_{T}}~,
\end{equation}
so that
\begin{equation}\label{e3-24}
A\equiv\sqrt{- 162 (\frac{E}{E_{Pl}})^{14} - 63 (\frac{E}{E_{Pl}})^{12} + 33 (\frac{E}{E_{Pl}})^{10}
+ 54 (\frac{E}{E_{Pl}})^8  + 83 (\frac{E}{E_{Pl}})^6 + 47 (\frac{E}{E_{Pl}})^4 + 8 (\frac{E}{E_{Pl}})^2
}~.
\end{equation}
The above solution shows that no matter how the energy levels of probe particle(s) increase the
interval between $a_{1}$ and $a_{2}$ becomes smaller up to where the oscillating universe
turns into a SU which is fluctuating around a non-zero value of the scale factor, see Figure
3 (right panel). Theoretically, this recent case is rather dramatic for two reasons. First, both of
the turning points correspond to non-zero values of the scale factor, in contrast to the initial
singular harmonic universe discussed in previous cases. Secondly, constraints (\ref{e3-22}) address
the existence of a positive cosmological constant which entails no violation of positive energy
requirement. At first sight, one may get the illusion that under conditions  (\ref{e3-22}) there is
an expanding and oscillating universe in the absence of the initial singularity. Expansion in rainbow
universe model in our study is due to the existence of a two  component source corresponding to repulsion,
that is $\omega<-\frac {1}{3}$ and $\Lambda>0$. Even though from a dynamical system point of view, the
minimum turning point is perfectly stable, Figure 3 (right panel) explicitly shows that in the QM view there
is a chance of the  minimum classical turning point collapsing by tunneling the potential barrier to zero
scale factor $a=0$. To be more specific, in Table 3 the outcome of a numerical analysis by means of
WKB-approximation is shown. These results can be interpreted as the probability of collapse via quantum
tunneling as the rainbow universe bounces at the scale factor $a_{1}$. Overall, this numerical analysis
reflects the fact that as the dimensionless ratio $\frac{E}{E_{Pl}}$ is getting close to unity, the chance
of the minimum scale factor collapsing becomes tiny and ignorable. Concretely speaking, the closed rainbow
universe model, when satisfying the constraints (\ref{e3-22}), can results in a non-singular oscillating
cosmology which at a high energy phase may become quantum mechanically stable.

\section{Quantization, wave function and initial singularity}
Let us now address the issue of initial singularity from the perspective of the solution of the
Schr\"{o}dinger-Wheeler-DeWitt (SWD) equation for the wave function of a closed rainbow universe. to this end, we start with the classical
super Hamiltonian (\ref{e2-11}) and try to quantize it. The existence of Lapse function $N$ signals the classical constraint
equation ${\cal H}=0$ because it explicitly refers to a Lagrange multiplier. So, the operator version of this constraint acting on
the wave function $\Psi(a,T)$ for a closed, $k=+1$, rainbow modified FRW model is written as
\begin{equation}\label{e4-1}
{\cal
H}\Psi(a,T)=\left(-\frac{P_a^2}{24a}
-\frac{6a}{f_{2}^{4}(\frac{E}{E_{Pl}})} + f_2^{3\omega-3}(\frac{E}{E_{Pl}})~ \frac{P_T}
{a^{3\omega}} - \frac{\Lambda a^{3}}
{f_2^{6}(\frac{E}{E_{Pl}})} \right)\Psi(a,T)=0 ~.
\end{equation}
Here, $\Psi(a,T)$ is the wave function of the universe.
We choose the ordering $a^{-1}P_a^2=P_aa^{-1}P_a$ to make the Hamiltonian Hermitian
and use the usual operator representations $(P_{a}, P_{T})\rightarrow -i(\frac{\partial}
{\partial_{a}},  \frac{\partial}{\partial_{T}})$ to find
\begin{eqnarray}\label{e4-2}
\left(\frac{1}{a}\frac{\partial^2\Psi(a,T)}{\partial a^2}-\frac{1}{a^{2}}\frac{\partial\Psi(a,T)}
{\partial a}-\frac{144a}{f_{2}^{4}(\frac{E}{E_{Pl}})}-\frac{24\Lambda a^{3}}{f_{2}^{6}(\frac{E}{E_{Pl}})}
-\frac{24iE}{a^{3\omega}}f_2^{3\omega-3}(\frac{E}{E_{Pl}})\frac{\partial\Psi(a,T)}{\partial T}\right)\Psi(a,T)=0~.
\end{eqnarray}
Introducing the following separation of variables
\begin{equation}\label{e4-3}
\Psi (a,T) = e^{-iET} \chi (a)~,
\end{equation}
Eq. (\ref{e4-2}) reduces to
\begin{eqnarray}\label{e4-4}
a\frac{\partial^2\chi(a)}{\partial a^2}-\frac{\partial\chi(a)}{\partial a}
-A_1a^3-A_2a^5+A_3a^{2-3\omega}\chi(a)=0
\end{eqnarray}
where
\begin{eqnarray}\label{e4-5}
A_1=144f_{2}^{-4}(\frac{E}{E_{Pl}}),~~~A_2=24\Lambda f_{2}^{-6}(\frac{E}{E_{Pl}})
,~~~A_3=24Ef_2^{3\omega-3}~.
\end{eqnarray}
The differential equation (\ref{e4-4}) in this form is not manageable. Given that
we are interested in the high energy (or short distance) phase of the universe, we expect
to drop terms of order $a^2$ or higher. Therefore, as can be seen, the term including  $\omega=-2/3$ in Eq. (\ref{e4-4}) does not contribute to the rainbow function
$f_2$. Thus, we look for the solution
and analysis of the differential equation relevant to the case of $\omega=0$, i.e.
\begin{eqnarray}\label{e4-5}
a\frac{\partial^2\chi(a)}{\partial a^2}-\frac{\partial\chi(a)}{\partial a}
+A_3a^{2}\chi(a)=0~.
\end{eqnarray}
Elimination of the terms  $a^3$ and $a^5$ in  equation (\ref{e4-4})
means that non flat spatial curvature and also non zero cosmological
constant corrections should not affect the wave function in the short distance regimes (around
the Planck scale). The general solution of  equation
(\ref{e4-5}) can be written in terms of Bessel functions $J_{\nu}$ and $Y_{\nu}$
\begin{eqnarray}\label{e4-8}
\chi(a)=a\left[C_{1}J_{2/3}\left(\sqrt{\frac{32E}{3}}
f_2^{\frac{3\omega-3}{2}}(\frac{E}{E_{Pl}})a^{3/2}\right)+C_2Y_{\nu}
\left(\sqrt{\frac{32E}{3}}
f_2^{\frac{3\omega-3}{2}}(\frac{E}{E_{Pl}})a^{3/2}\right)\right].
\end{eqnarray}
Here, $C_{1,2}$ can be interpreted as integration constants. However, we should set
$C_{2}=0$ in the above solution since  $Y_{\nu}$ goes to infinity at the origin. Thereupon,
the final form of the eigenfunction of SWD equation reads
\begin{eqnarray}\label{e4-9}
\psi(a,T)=e^{-iET}a\left[J_{2/3}\left(\sqrt{\frac{32E}{3}}
f_2^{\frac{3\omega-3}{2}}(\frac{E}{E_{Pl}})a^{3/2}\right)\right].
\end{eqnarray}
Let us now introduce a weight function  $A(E)$ and write
the wave packet solution to the SWD equation as
\begin{equation}\label{e4-10}
\Psi(x,T)=\int_{E=0}^{\infty} A(E)\Psi_{E}(x,T) dE~.
\end{equation}
In order to come to an analytical expressions for the integral relation
(\ref{e4-10}) we introduce the following quasi-Gaussian weight factor
for the function $A(E)$
\begin{equation}\label{e4-11}
A(E)=\left(\frac{32E}{3}\right)^{1/3}\exp\left(-\frac{32\gamma
E}{3}\right)~,
\end{equation}
where $\gamma$ is a positive numerical factor. Now, Eq. (\ref{e4-10}) reads
\begin{equation}\label{e4-12}
\frac{16a}{3} \int_{0}^{\infty} \left(\frac{32E}{3}\right)^{1/3}\exp\left
(-\frac{32\gamma E}{3}\right)e^{-iET}\left[J_{2/3}\left(\sqrt{\frac{32E}{3}}
f_2^{-\frac{3}{2}}(\frac{E}{E_{Pl}})a^{3/2}\right)\right] dE~.
\end{equation}
Setting $\eta=\sqrt{\frac{32E}{3}}$, the above integral becomes
\begin{equation}\label{e4-13}
a\int_{0}^{\infty}\exp\left(-\alpha\eta^{2}\right)\eta^{5/2}J_{2/3}\left(\eta
f_2^{-\frac{3}{2}}(\frac{E}{E_{Pl}})a^{3/2}\right)d\eta,
\end{equation}
with $\alpha=(\gamma+\frac{3i}{32}T)$. The final step before solving is to noted that
due to the existence of an explicit cutoff in the energy scale
at which the minisuperspace  is probed by test photons, the measure of the
integral under consideration over $E$ is deformed as $dE \rightarrow
f_2(\frac{E}{E_{Pl}})dE$. Therefore, by including the rainbow functions (\ref{e3-5})
and (\ref{e3-6}), the integral (\ref{e4-13}) takes the form
\begin{equation}\label{e4-14}
a\int_{0}^{\infty}\exp\left(-\alpha\eta^{2}\right)\eta^{5/2}J_{2/3}\left(\eta
a^{3/2}\left(1-\frac{9\eta^{2}}{64E_{Pl}}\right)\right)\left(1+\frac{3\eta^{2}}{32E_{Pl}}
\right)d\eta,
\end{equation}
and
\begin{equation}\label{e4-15}
a\int_{0}^{\infty}\exp\left(-\alpha\eta^{2}\right)\eta^{5/2}J_{2/3}\left(\eta
a^{3/2}\left(1+\frac{3\eta^{2n}}{128(E_{Pl})^{n}}\right)\right)
\left(1-\frac{3\eta^{2n}}
{64(E_{Pl})^{n}}\right)d\eta~,
\end{equation}
respectively.
Finally, we get the following expressions
\begin{eqnarray}\label{e4-16}
\psi(a,T)&=&{\cal{N}}\left[a^{2}\left(2\gamma+\frac{3i}{16}T\right)^{-5/3}+\frac{3}{32E_{Pl}}
\left(2\gamma+\frac{3i}{16}T\right)^{-11/3}\times\right. \nonumber\\
& &\left.
\left(a^{5}+\left(\frac{20\gamma}{3}+\frac{5i}
{8}T\right)a^{2}\right)\right]
\exp\left(-\frac{a^3}{\left(4\gamma+\frac{3i}{8}T\right)}\right)
\end{eqnarray}
and
\begin{eqnarray}\label{e4-17}
\psi(a,T)&=&{\cal{N}}\left[a^{2}\left(2\gamma+\frac{3i}{16}T\right)^{-5/3}
-\frac{9}{2048E_{Pl}^{2}}\left(2\gamma+\frac{3i}{16}T\right)^{-17/3}\times
\right. \nonumber\\
& &\left.
\left(a^8-\frac{64}{3}\left(\gamma+\frac{3i}{32}T\right)a^5 +\frac{640}{9}
\left(\gamma+\frac{3i}{32}T\right)^{2}a^2\right)\right]
\exp\left(-\frac{a^3}{\left(4\gamma+\frac{3i}{8}T\right)}\right)
\end{eqnarray}
for the wave function. Note that we have used (see \cite{Abr})
$$\int_0^\infty e^{-a x^2}z^{\nu+1}J_{\nu}(bx)dx=\frac{b^{\nu}}{(2a)^{\nu+1}}e^{-\frac{b^2}{4a}},$$
$$\int_0^\infty e^{-a x^2}z^{\nu+5}J_{\nu}(bx)dx=\frac{b^{\nu}}{(2a)^{\nu+5}}e^{-\frac{b^2}{4a}}
\left(b^{4}-8(\nu+2)ab^{2}+16(\nu+1)(\nu+2)a^{2}\right).$$ We have also applied the approximations
$\left(1-\frac
{9\eta^{2}}{64E_{Pl}}\right)~\approx 1$ and $\left(1+\frac{3\eta^{2n}}{128(E_{Pl})^{n}}\right) ~\approx 1$
in the argument of the Bessel functions. Here, ${\cal{N}}$ is a numerical factor used for normalization purposes.
Our analysis in the previous section has shown that the values of $n$ are limited to even numbers since
to obtain the wave function (\ref{e4-17}) we have set $n=2$. At first glance, one notices that
these wave functions (\ref{e4-16}) and (\ref{e4-17}) go to zero as one approaches the origin (i.e. $a\rightarrow0$).
This means that non zero bouncing rainbow universe addressed in the previous section is free of quantum
collapse, recalling that the boundary condition $\psi(a\rightarrow0,T\rightarrow0)=0$ was first suggested in
\cite{W}.  Figure 4 shows the behavior of probability density $|\Psi(a, T)|^2$ using (\ref{e4-16}). We see that
the wave packet is peaked at the non-zero
values of $a$ at $T=0$ which supports the idea of non zero bouncing universe. We also find that increasing the value
of $\gamma$, the peak of $|\Psi(a, T)|^2$ emerges in higher non zero values of $a$ at $T=0$. Needless to say that generally all the results for the wave packet of the rainbow universe (\ref{e4-17}) are retained here as well. As final word in the section, we recommended reference
\cite{RE} in which displayed one of other benefits of the WDW equation within gravity's rainbow proposal. Briefly, in
the mentioned reference using a WDW equation (within a FRW and spherically symmetric background, respectively)
authors can show the equivalence of gravity's rainbow  with Ho\v{r}ava-Lifshitz model of gravity as theories which incorporate a
correction in the high energy regime.
\begin{figure}[htp]
\begin{center}\includegraphics{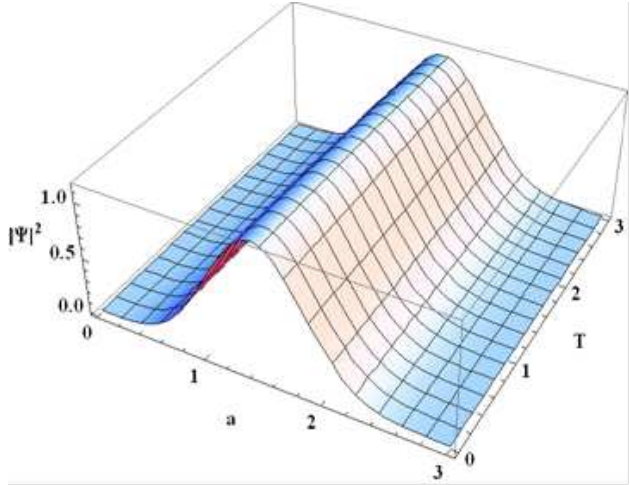} \vspace{6cm}
\end{center}
\caption{\small {The probability density function $|\Psi(a,T)|^2$
derived from the wave packet (\ref{e4-16}). To plot this figure
we have set the numerical values $\gamma = 1=E_{Pl}$ and ${\cal{N}}=8$.
Here, the peak of $|\Psi(a,T)|^2$ emerges around $a=1$ at $T=0$.}}
\label{fig:w}
\end{figure}

\section{Conclusions}
Following the previous studies, in this work  we use ``rainbow metric''
(\ref{e2-2}) as one of many possible hypotheses to provide an effective
description of some QG effects in the early stages of the formation of
the universe. Our principal goal in this paper was dedicated to the
investigation of the initial singularity issue at the quantum level
in a closed rainbow cosmology with a homogeneous isotropic space-time
background. We used a one-dimensional minisuperspace model including
the dimensionless ratio $\frac{E}{E_{Pl}}$, for an ideal fluid with a
cosmological constant and constructed the Hamiltonian, in the framework
of the Schutz's formalism.  It should be stressed that the dimensionless
ratio $\frac{E}{E_{Pl}}$ directly arises from the rainbow gravity proposal.
For the matter sources we considered the following cases:
\begin{enumerate}
  \item Non-relativistic dust matter with $\omega=0$ plus the cosmological
  constant $\Lambda$,
  \item Exotic matter as domain wall fluid with $\omega=-\frac{2}{3}$ plus
 the cosmological constant $\Lambda$.
\end{enumerate}
For each of these two cases, our closed FRW universe model was modified by
rainbow functions (\ref{e3-5}) and (\ref{e3-6}). By analyzing the potential
sector of the relevant Hamiltonian, we demonstrated that for both options
above, when certain conditions are satisfied, we get either a potential
barrier or a static universe for the non-zero values of the scale factor.
These two situations provide the possibility of removing the initial singularity.
However, the main point is that if the non-zero values of the scale factor
become unstable, any discussion of the initial singularity becomes meaningless
and misleading. As we saw, from a classical dynamical system viewpoint, they
can be stable. This does not rule out the non-zero probability from a QM
perspective  of a return to a zero value scale factor via tunneling. By considering
the shape of the potential as well as using an approximation procedure such
as WKB, we showed that the higher the energy levels of probe particle(s)
(i.e. $\frac{E}{E_{Pl}}$), the higher the barrier height. That is, the higher
the probing energy, the chance of tunneling through the barrier to a zero value
of the scale factor diminishing to zero. Therefore, despite the probability
of minimal scale factor crumbling in both closed rainbow universe models,
for the energy-dependent space-time geometry with energies around the Planck
energy, there is the possibility of quantum stability.

We also noted that the rainbow function (\ref{e3-6}) modification to FRW universe
with a domain wall fluid plus cosmological constant provided the constraints
(\ref{e3-22}), which in turn resulted in a non-singular harmonic universe.  The
behavior of this harmonic universe is interesting in the sense that it is oscillating
between minimum and maximum values of the scale factor and is sensitive to the
dimensionless ratio $\frac{E}{E_{Pl}}$ (Eq. (\ref{e3-23})).  Figure 3 (right panel) displays
this behavior in that by increasing the value of $\frac{E}{E_{Pl}}$, the oscillating interval
becomes smaller and smaller in such a way that at the high energy phase the harmonic
universe reduces to a SU. We also demonstrated that such a harmonic universe could be
stable quantum mechanically at high energies.

Finally, we quantized
the classical Hamiltonian (\ref{e2-13})
within the rainbow framework and used separation of variable method to obtain
 analytical solutions relevant to the SWD equation. Introduction of
a suitable superposition of the eigenfunctions was then used to derive the wave packet
of the universe modified by two rainbow functions $f_2$, (\ref{e3-5}) and $f_1$, (\ref{e3-6})
, respectively. We showed that both wave functions
(\ref{e4-16}) and (\ref{e4-17}) satisfy the boundary condition $\psi(a\rightarrow0,T\rightarrow0)$. This would mean that  the non zero bouncing rainbow universe  will not lead to a quantum collapse. The peak of the probability density function $|\Psi(a,T)|^2$, shown in Figure 4, coincides with the non zero value of $a$ at $T=0$ which
supports a bouncing non-singular universe.


\begin{thebibliography}{00}
\bibitem{Geo}
G. Lemaitre, Gen. Rel. Grav. 29 (1997) 641
\bibitem{Pen}
R. Penrose. Phys. Rev. Lett. 14(3)  (1965) 57
\bibitem{SW0}
S. W. Hawking. P. Roy. Soc. A-Math. Phy. 294(1439) (1966) 511
\bibitem{SW1}
S. W. Hawking. P. Roy. Soc. A-Math. Phy. 295(1443) (1966) 490
\bibitem{SW2}
S. W. Hawking. P. Roy. Soc. A-Math. Phy. 300(1461) (1967) 187
\bibitem{PH}
S. W. Hawking and G. F. R. Ellis, ``The Large Scale Structure of
Space-Time'', Cambridge University Press, (1973)
\bibitem{J}
J. Earman, ``Bangs, Crunches, Whimpers and Shrieks: Singularities
and A causalities in Relativistic Spacetimes'', Oxford University
Press, USA (1995)
\bibitem{A0}
D. Battefeld and P. Peter, Phys. Rept. 571 (2015) 1
\bibitem{A1}
L. J. Garay, M. Martin-Benito and E. Martin-Martinez, Phys. Rev.
D 89 (2014) 043510
\bibitem{A2}
N. Pinto-Neto and J. C. Fabris, Class. Quant. Grav. 30 (2013)  143001
\bibitem{A3}
A. Ashtekar and P. Singh, Class. Quant. Grav. 28  (2011) 213001
\bibitem{A4}
R. Brandenberger, Phys. Rev. D 80  (2009) 043516
\bibitem{A5}
G. Calcagni, JHEP 0909 (2009) 1112
\bibitem{LQ0}
R. Gambini, and J. Pullin, Phys. Rev. D 59 (1999) 124021
\bibitem{LQ1}
A. Ashtekar, J. Lewandowski, Class. Quant. Grav. 21, R53-R152
(2004)
\bibitem{St}
V. A. Kosteleck´y, and S. Samuel, Phys. Rev. D 39 (1989) 683
\bibitem{Ma0}
M. Bojowald.  Phys. Rev. Lett. 86 (2001) 5227
\bibitem{Ma1}
A. Ashtekar, T. Pawlowski, P. Singh, Phys. Rev. Lett.
96 (2006) 141301
\bibitem{Ma2}
A. Ashtekar, T. Pawlowski, P. Singh,  Phys. Rev. D 74
(2006) 084003
\bibitem{Ben}
B. Craps, T. Hertog, and N. Turok. Phys.
Rev. D 80  (2009) 086007
\bibitem{GUP0}
A. Kempf, G. Mangano and R. B. Mann, Phys. Rev. D 52 (1995) 1108
\bibitem{GUP1}
R. J. Adler, P. Chen and D. I. Santiago, Gen. Rel. Grav. 33 (2001) 2101
\bibitem{GUP2}
A. J. M. Medved and E. C. Vagenas, Phys. Rev. D 70 (2004) 124021
\bibitem{GUP3}
Y. Ling, B. Hu and X. Li, Phys. Rev. D 73  (2006) 087702
\bibitem{GUP4}
B. Vakili, Phys. Rev. D 77 (2008) 044023
\bibitem{GUP5}
A. F. Ali, S. Das and E. C. Vagenas, Phys. Lett. B 678 (2009) 497
\bibitem{GUP6}
S. Das, E. C. Vagenas and A. F. Ali, Phys. Lett. B 690 (2010) 407
\bibitem{GUP7}
K. Nozari and S. Saghafi, JHEP 1211 (2012) 005
\bibitem{GUP8}
K. Nozari and  A. Etemadi, Phys. Rev. D 85 (2012) 104029
\bibitem{GUP9}
S. Jalalzadeh, M. A. Gorji and K. Nozari, Gen. Rel. Grav. 46 (2014)
1632
\bibitem{GUP10}
P. Pedram, K. Nozari and S. H. Taheri, JHEP 03 (2011) 093
\bibitem{GUP11}
K. Nozari, M. Khodadi, M. A. Gorji, Europhys. Lett. 112 (2015) 60003
\bibitem{GUP12}
K. Nozari, M. A. Gorji, V. Hosseinzadeh and B. Vakili, Class. Quantum
Grav. 33 (2016) 025009
\bibitem{Ame0}
G. Amelino-Camelia, Phys. Lett. B 510 (2001) 255
\bibitem{Ame1}
G. Amelino-Camelia, Int. J. Mod. Phys. D 11 (2002) 35
\bibitem{Mag0}
G. Amelino-Camelia, J. Kow
alski-Glikman, G. Mandanici
and A. Procaccini, Int. J. Mod. Phys. A 20 (2005) 6007
\bibitem{Mag1}
J. Magueijo and L. Smolin, Phys. Rev. D 67 (2003) 044017
\bibitem{Mag2}
J. Magueijo and L. Smolin, Phys. Rev. Lett. 88 (2002) 190403
\bibitem{Mag3}
J. Magueijo and L. Smolin, Class. Quant. Grav. 21 (2004) 1725
\bibitem{Ali}
A. Awad, A. F. Ali, B. Majumder, JCAP 10  (2013) 052
\bibitem{Bar}
B. Majumder, Int. J. Mod. Phys. D 22 (2013) 1350079
\bibitem{Vilen0}
A. T. Mithani, A. Vilenkin, JCAP 1201 (2012) 028
\bibitem{Vilen1}
A. Borde, A. H. Guth and A. Vilenkin, Phys. Rev. Lett.
90 (2003) 151301
\bibitem{Talk0}
S. del Campo, E. Guendelman, A. B. Kaganovich, R. Herrera
and P. Labrana, Phys. Lett. B 699 (2011) 211
\bibitem{Talk1}
P. Wu and H. Yu, Phys. Rev. D  81 (2010) 103522
\bibitem{Talk2}
D. J. Mulryne, R. Tavakol, J. E. Lidsey and G. F. R. Ellis,
Phys. Rev. D 71 (2005) 123512
\bibitem{Ellis1}
G. F. R. Ellis and R. Maartens, Class. Quant. Grav. 21 (2004) 223
\bibitem{Ellis2}
G. F. R. Ellis, J. Murugan and C. G. Tsagas, Class. Quant. Grav. 21
(2004) 233
\bibitem{CMB0}
C. L. Bennet, et al., Astrophys. J. Suppl. 148 (2003) 1
\bibitem{CMB1}
D. N. Spergel, et al.,  Astrophys. J. Suppl. 148 (2003) 175
\bibitem{Mohsen1}
M. Khodadi, Y. Heydarzade, K. Nozari, F. Darabi,
Eur. Phys. J. C 75 (2015) 590
\bibitem{sch0}
B. F. Schutz, Phys. Rev. D 2 (1970) 2762
\bibitem{sch1}
B. F. Schutz, Phys. Rev. D 4  (1971) 3559
\bibitem{Lap}
V. G. Lapchinskii and V. A. Rubakov, Theor. Math. Phys. 33
(1977) 1076
\bibitem{Babak}
B. Vakili, Phys. Lett. B  688 (2010) 129
\bibitem{Mohsen2}
M. Khodadi, K. Nozari, B. Vakili, Gen. Rel. Grav. 48 (2016) 64
\bibitem{Li}
Yi Ling and Qingzhang Wu, Phys. Lett. B 687 (2010) 103
\bibitem{Amelino1}
G. Amelino-Camelia, J. R. Eliss, N. Mavromatos and D. V.
Nanopoulos, Int. J. Mod. Phy. A. 12 (1997) 607
\bibitem{PI}
G. Amelino-Camelia, Living Rev. Rel. 16 (2013) 5
\bibitem{Dab}
M. P. Dabrowski and A. L. Larsen, Phys. Rev. D  52 (1995) 3424
\bibitem{Atk}
D. Atkatz, Am. J. Phys. 62 (1994) 619
\bibitem{BUC}
M. Bucher and D. N. Spergal, Phys. Rev. D 60 (1999)
043505
\bibitem{1974}
Y. B. Zeldovich, I. Y. Kobzarev, and L. B. Okun, Zh.
Eksp. Teor. Fiz. 67 (1974) 3 (also in Sov. Phys. JETP, 40 (1974) 1)
\bibitem{Abr}
M. Abramowitz and I. A. Stegun, ``Handbook of Mathematical Functions'', New York: Dover (1972)
\bibitem{W}
B. S. Dewitt, Phys. Rev. 160 (1967) 1113
\bibitem{RE}
R. Garattini and E. N. Saridakisc, Eur. Phys. J. C 75 (2015) 343
\end{thebibliography}
\end{document}